\documentclass[12pt]{article}
\usepackage{a4}
\usepackage{amsmath}
\usepackage{amsfonts}     
\usepackage{epsfig}
\def\pa{\partial}

\def\nn{\nonumber}
\def\ga{\gamma}

\def\si{\sigma}

\def\de{\delta}

\def\al{\alpha}

\def\la{\lambda}
\def\La{\Lambda}
\def\eps{\epsilon}

\def\dd{{\rm d}}
\def\cL{{\cal L}}
\def\cD{{\cal D}}

\def\vD{{\vec D}}

\newcommand{\ZZ}{\mathbb{Z}}

\begin{document}

\begin{flushright} 
hep-th/0203055
\end{flushright} 
\vspace{20mm}
\begin{center}
{\Large {\bf Spontaneous localization of bulk matter fields}
}
\\[0pt]
\bigskip
\bigskip {\large
{\bf S.\ Groot Nibbelink ${}^{a,}$\footnote{
{{ {\ {\ {\ E-mail: nibblink@th.physik.uni-bonn.de}}}}}}}, 
{\bf H.P.\ Nilles ${}^{a,}$\footnote{
{{ {\ {\ {\ E-mail: nilles@th.physik.uni-bonn.de}}}}}}},
{\bf M.\ Olechowski ${}^{a,b,}$\footnote{
{{ {\ {\ {\ E-mail: olech@th.physik.uni-bonn.de}}}}}}}
\bigskip }\\[0pt]
\vspace{0.23cm}
{\it ${}^a$ Physikalisches Institut der Universit\"at Bonn,} \\
{\it Nussallee 12, 53115 Bonn, Germany.}\\
\vspace{0.23cm}
{\it ${}^b$ Institute of Theoretical Physics, Warsaw University,} \\
{\it Ho\.za 69, 00--681 Warsaw, Poland.}\\

\bigskip
\vspace{30mm}
Abstract
\end{center}

\noindent
We study models compactified on $S^1/\ZZ_2$ with bulk and brane
matter fields charged under $U(1)$ gauge symmetry. We calculate
the FI--terms and show by minimizing the resulting potential that
supersymmetry or gauge symmetry is spontaneously broken if the sum of
the charges does not vanish. Even if this sum vanishes, there could 
be an instability as a consequence of localized FI--terms. This leads
to a spontaneous localization of charged bulk fields on respective
branes.

\newpage
Theories in more than 4--dimensional space--time have
attracted much attention in recent years. The connection to our 
4--dimensional world is made through compactification of
extra space dimensions. Among the proposed mechanisms,
orbifold compactification seems to be particularly simple
and instructive \cite{dixon}. In addition, it allows the
incorporation of so--called brane world scenarios, in which
branes are placed (at fixed points)
in a higher dimensional bulk. We  can then 
distinguish between bulk--fields that propagate in the higher
dimensional bulk and brane fields whose propagation and 
interactions are confined to the respective branes (fixed points).

Various models of this kind have been proposed recently because of
their potential phenomenological merits. 
Some unwanted couplings can be made very small or even zero
by appropriate localization of the fields. For example, two fields
localized at different branes can not interact directly but only
indirectly via bulk fields. 
In addition,
two such brane fields could have 
different couplings to a bulk field if that field has nontrivial
profile along the extra dimensions. This type of mechanism could
be responsible for the suppression of proton decay, Yukawa
couplings and supersymmetry breakdown.

Many of these models assume some amount of supersymmetry to 
render the theory ultraviolet (UV) insensitive and solve 
the hierarchy problem. 
While in 4 dimensions the UV--behavior of the theory is well 
understood there remain some open questions in the brane world
picture. It is one of these questions that we shall study in the
present paper: the UV--sensitivity of the low-energy physics.

In 4 dimensions the supersymmetric models (also with soft
breaking interactions) are usually at most logarithmically sensitive
to the high scale $\La$. The only exception 
(the Achilles' heel of supersymmetry) is the quadratically
divergent Fayet--Iliopoulos term for a $U(1)$ gauge symmetry 
if the sum over the scalar fields of the corresponding charge 
is nonzero \cite{fischler}. 
Thus, in phenomenologically acceptable models the $U(1)$ charges must
sum up to zero; otherwise the low energy physics will be destabilized 
and either gauge symmetry or supersymmetry will be broken at a
scale of order $\La$. Another (independent) reason for this condition 
is the absence of the mixed gauge--gravitational anomaly if 
gravity is taken into account. 

The Standard Model of strong and electroweak interactions 
contains a $U(1)$ gauge symmetry -- hypercharge symmetry. 
The sum of hypercharges vanishes for each
generation of fermions so also for each generation of scalars in the
supersymmetric version of the model. Thus, there are no 
quadratically divergent FI--terms in
the 4--dimensional supersymmetric extension of the standard model.

We would now like to ask the question how this situation generalizes
to the higher dimensional brane world scenario. 
Previously \cite{GGNN} we have already shown that the presence of
massless 
bulk scalar fields with nonvanishing sum of charges leads to a
quadratic  
divergence\footnote{For a related discussion of ultraviolet
sensitivity in extra dimensions see \cite{GN,G,GNS}.}. 
In this paper we  
set up the general case with charged bulk and brane fields and
arrive at a similar conclusion: if the sum of charges does not
add up to zero, there is a quadratically divergent FI--term and
either supersymmetry or gauge symmetry are destabilized in the
same way as in the 4--dimensional theory.

Still this is not the full story. We next concentrate on the more
specific question of localization of the field with nonzero
hypercharge. Usually in brane world models some of the standard model
particles can be described by bulk fields, while others by brane
fields. 
Thus generically the hypercharges add up to zero only {\bf globally}
(when integrated over the extra dimension) but not {\bf locally}.
In such a situation we can show that as a result of the localized
FI--terms we find a {\bf spontaneous localization} of charged bulk
fields. 
The zero modes of these bulk fields become localized on the
brane, while the masses of the 
higher Kaluza-Klein modes are pushed to the UV--scale $\La$.

In the present paper we shall present our results in the framework
of the simplest possible toy model. The general results for more
realistic  
models will be presented in a future publication \cite{future}.
We consider the 5--dimensional supersymmetric model compactified on
$S^1/\ZZ_2$ containing 3 multiplets: two 5--dimensional bulk
multiplets -- the vector multiplet $V$ and the hypermultiplet $H$ --
and one 4--dimensional chiral multiplet $C_0$ localized at the brane
at $y=0$. One could expect a gauge anomaly in such a model 
similar to the one discussed in \cite{SSSZ}. However we will 
postpone the discussion of the gauge anomalies here for two reasons. 
First, the simple model allows a simplest possible illustration 
of the mechanism of localization observed here, and secondly, the 
localization of fields might actually lead to a more refined and
subtle discussion of gauge anomalies, including a possible
cancellation mechanism via Chern-Simons terms \cite{PR}.

The 5--dimensional off--shell vector multiplet 
$V=(A_M,\la^i, \Phi, \vD)$ contains a gauge field $A_M$, a
doublet of symplectic--Majorana gauginos $\la^i$, a real scalar $\Phi$
and a triplet of auxiliary fields $\vD$ 
(we use the standard notation: $M=1\ldots 5$, $\mu=1\ldots 4$, $x^5=y$).
Each of the components
must have a definite parity under the $\ZZ_2$ symmetry used for
orbifolding. We choose the parity assignments as follows:
\[
V:~
\begin{array}{|l|c|c|c|c|c|c|}
\hline
\text{state} & A_\mu & A_5 & \Phi & \la_\pm & D^3 & D^{1,2}
\\\hline 
\text{parity} & + & - & - & \pm & + & -
\\\hline
\end{array}
\]
With this choice neither the $N=1$ supersymmetry nor
the $U(1)$ gauge symmetry is broken by orbifolding.

The 5--dimensional off--shell hypermultiplet $H$ consists of two
complex scalars $\phi_\pm$, one Dirac fermion $\psi$, 
and two complex auxiliary fields $F_\pm$. The subscripts on the
bosonic components denote their parity under $\ZZ_2$:
$\phi_\pm(-y)=\pm\phi_\pm(y)$, $F_\pm(-y)=\pm F_\pm(y)$. The parity
of the components of $\psi$ is related  
to their 4--dimensional chirality: $\psi(-y)=i\ga^5\psi(y)$.

The bulk interactions of $V$ and $H$ are described by the standard
5--dimensional supersymmetric Lagrangian. The part of that Lagrangian
which is important for our discussion can be written, in terms of the
above introduced fields, as:
\begin{eqnarray}
\cL_{\rm bulk}
=
\!\!\!&-&\!\!\!
\frac12 \left(\pa_M\Phi\right)^2 + \frac12 \vD^2
-\cD_M\phi_+^\dagger\cD^M\phi_+ - \cD_M\phi_-^\dagger\cD^M\phi_-
+i\bar\psi\ga^M\cD_M\psi
\nn\\
\!\!\!&-&\!\!\!
gqD^3\left(\phi_+^\dagger\phi_+ - \phi_-^\dagger\phi_-\right)
-gq\left(\left(D^1-iD^2\right)\phi_+^T\phi_- +{\rm h.c.}\right)
\nn\\
\!\!\!&-&\!\!\!
g^2q^2\Phi^2\left(\phi_+^\dagger\phi_+ + \phi_-^\dagger\phi_-\right)
-gq\Phi\bar\psi\psi
+\ldots
\,.
\label{Lbulk}
\end{eqnarray}

Now we add a 4--dimensional chiral multiplet 
$C_0=(\phi_0, \psi_0, F_0)$ localized at the brane at $y=0$. 
It was shown in ref.\ \cite{MP} that such a brane multiplet can
be coupled to the bulk gauge multiplet in a way which preserves $N=1$ 
supersymmetry. The coupling is given by the standard 4--dimensional
interaction of a chiral multiplet with a gauge multiplet when the role
of a 4--dimensional gauge multiplet is played by the boundary values
of the appropriate components of the gauge bulk 
field\footnote{
In a similar way the brane chiral multiplet can be coupled to the
chiral multiplet made out of the boundary values of the bulk
hypermultiplet components ($\phi_+, \psi_+, F_+ - \pa_y\phi_-$). More
details will be presented in a future publication \cite{future}.
}: 
$(A_\mu, \la_+, D^3-\pa_y\Phi)$. The resulting Lagrangian reads
\begin{equation}
\cL_{\rm brane}
=
\de(y)\left(
-\cD_\mu\phi_0^\dagger\cD^\mu\phi_0 
+ gq_0\left(D^3-\pa_y\Phi\right)\phi_0^\dagger\phi_0
+\ldots
\right)
\label{Lbrane}
\end{equation}
where $q_0$ is the $U(1)$ charge of $C_0$.
The tree level action of the model is given by the sum of the bulk and
brane contributions
\begin{equation}
S^{(0)}
=
\int\dd^4x\dd y 
\big(\cL_{\rm bulk} + \cL_{\rm brane}\big)
\label{action}
\end{equation}
and is invariant under the $N=1$ supersymmetry and the $U(1)$ gauge
symmetry.

Let us now take the 1--loop corrections into account. It occurs that 
divergent FI--terms play a very crucial role. They 
are generated by tadpole diagrams even if the sum of all charges is
zero because the charge is non--trivially distributed along the fifth
dimension.

We start with the contribution coming from the tadpole with the
brane scalar $\phi_0$ in the loop (fig.\ 1a). It is given by the
standard 4--dimensional result with the gauge auxiliary field replaced
by the boundary value of the combination ($D^3 - \pa_y\Phi$). In the
cut--off regularization it is given by
\begin{equation}
\left(D^3 - \pa_y\Phi\right)
gq_0\frac{\La^2}{16\pi^2}\de(y)
\,.
\label{FIbrane}
\end{equation}

The calculation of the bulk field contributions is more involved 
(part of it has been already presented in refs.\ \cite{GGNN, SSSZ}).
It can be performed by expanding the bulk fields into 4--dimensional
modes and summing the contributions from the obtained tadpole
diagrams. The two kinds of bulk tadpole diagrams are shown on 
figs.\ 1b and 1c.
In one of them the $D^3$ field is coupled to the scalar loops and only 
this contribution has been taken into account in the previous
calculations \cite{GGNN, SSSZ}. But there is also a contribution in
which the $\Phi$ scalar is coupled to the fermion loop. 
It gives the same FI--term 
for $(-\pa_y\Phi)$ as the bulk scalar tadpole gives for $D^3$. 
One should expect this because the model
is invariant under $N=1$ supersymmetry and the combination
($D^3-\pa_y\Phi$) is the gauge auxiliary field under this
supersymmetry \cite{MP}. 
The details of
the calculation will be presented elsewhere \cite{future} and here we
give only the final result:
\begin{equation}
\left(D^3-\pa_y\Phi\right)
g\frac{q}{2}
\left\{
\frac{\La^2}{16\pi^2}
\left[\de(y)+\de(y-\pi R)\right]
+\frac{\ln\La^2}{16\pi^2}\left[\de''(y)+\de''(y-\pi R)\right]
+\ldots
\right\}
\label{FIbulk}
\end{equation}
where dots denote the finite contribution which will not be discussed 
here because we are interested only in the sensitivity of the model to
the high energy scale represented by the cut--off $\La$. 
%
%

Let us now analyse our model with those 1--loop corrections 
quadratically and logarithmically sensitive to the cut--off scale
$\La$. We start with the 4--dimensional effective potential which is
given by 
the sum of the potential part of the tree level action (\ref{action})
and the integrals of (\ref{FIbrane}) and (\ref{FIbulk}).
The part of the potential important for our analysis (i.e. without the
$F$-type terms) reads
\begin{eqnarray}
V
=
\int{\rm d}y
&&\!\!\!\!\!\!\!\!\!
\Big[
-\frac12\vD^2+\frac12\left(\pa_y\Phi\right)^2
+\left|\pa_y\phi_+\right|^2+\left|\pa_y\phi_-\right|^2
+\xi\left(D^3-\pa_y\Phi\right)
\nn\\
&&\!\!\!\!\!\!
+gq_0\left(D^3-\pa_y\Phi\right)\left|\phi_0\right|^2\de(y)
+g^2q^2\Phi^2\left(\left|\phi_+\right|^2+\left|\phi_-\right|^2\right) 
\nn\\
&&\!\!\!\!\!\!
+gqD^3\left(\left|\phi_+\right|^2-\left|\phi_-\right|^2\right)
+gq\left(\left(D^1-iD^2\right)\phi_+ \phi_- +{\rm h.c.}\right)
\Big]
\end{eqnarray}
where the divergent radiatively generated FI--parameter $\xi(y)$ 
is given by
\begin{eqnarray}
\xi
\!\!\!&=&\!\!\!
g\frac{\La^2}{16\pi^2}\!\!
\left[\left(q_0+\frac{q}{2}\right)\de(y)+\frac q2 \de(y\!-\!\pi R)\right]
+g\frac{\ln\La^2}{16\pi^2}\!\!
\left[\frac{q}{2}\de''(y)+\frac{q}{2}\de''(y\!-\!\pi R)\right].
\,\,
\label{xi}
\end{eqnarray}
We have dropped the $A_5$ component of the gauge field: as $A_5$
vanishes at both boundaries, we can perform a gauge transformation to put it
zero over the bulk. 

\begin{figure}
\begin{center}
\scalebox{1.0}{\mbox{\input{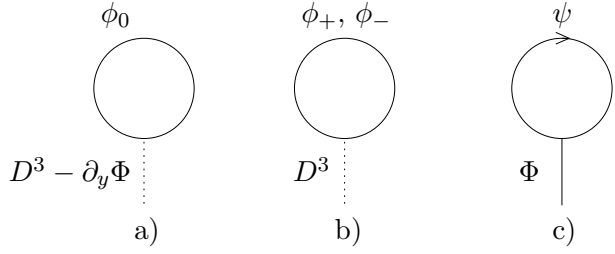}}}
\end{center}
\caption{The diagrams that give rise to FI--terms 
via brane scalar (a), bulk scalar (b) and bulk fermion (c) loops.
}
\end{figure}

We rewrite this potential in a form more suitable for our later
discussion. We add and subtract the term 
$gq(\pa_y\Phi)
\left(\left|\phi_+\right|^2-\left|\phi_-\right|^2\right)$ 
and perform an integration by parts, 
to make it possible to write the potential as a combination of
squares: 
\begin{eqnarray}
V 
=
\int\dd y
\Big[
\!\!\!&&\!\!\! \!\! 
-\frac12\left( - D^3 + 
\xi+gq_0\left|\phi_0\right|^2\de(y)
+gq\left|\phi_+\right|^2-gq\left|\phi_-\right|^2\right)^2
\nn\\
&&\!\!\! \!\! 
+\frac12\left( -\pa_y\Phi + 
\xi+gq_0\left|\phi_0\right|^2\de(y)
+gq\left|\phi_+\right|^2-gq\left|\phi_-\right|^2
\right)^2
\nn\\
&&\!\!\! \!\! 
- \frac 12 \left| D^1 + i D^2 +  gq \phi_+ \phi_- \right|^2 
+  g^2 q^2  \left| \phi_+ \phi_- \right|^2 
\nn\\
&&\!\!\! \!\! 
+ \left|\pa_y\phi_+-gq\Phi\phi_+\right|^2
+\left|\pa_y\phi_-+gq\Phi\phi_-\right|^2
\Big].
\label{V}
\end{eqnarray}
By splitting of the squares, it seems that we have introduced squares
of delta functions or their derivatives. However, because of the opposite
sign of the auxiliary field $D^3$ all this cancels out
precisely. Thus the dangerous squares of
$\de(y)$ and $\de''(y)$ do not appear. (Here we presented only this 
simplified argumentation and we postpone a more complete calculation
to a future publication \cite{future}).
As the only negative terms in this potential are due to the
auxiliary fields, that have algebraic equations of motion, 
for example 
\begin{equation}
D^3(y)
=
\xi(y)+gq_0\left|\phi_0\right|^2\de(y)
+gq\left|\phi_+(y)\right|^2-gq\left|\phi_-(y)\right|^2
\,,
\label{D3}
\end{equation}
the potential \eqref{V} is clearly positive semi--definite.

Next, we investigate background field
configurations, i.e.\ Vacuum Expectation Values (VEVs) of the scalar
fields (which may be functions of $y$) that minimize the potential $V$. 
An important question is whether 
supersymmetry or $U(1)$ gauge symmetry is spontaneously
broken. Gauge symmetry remains unbroken if the charged fields have 
vanishing VEVs: 
$\left<\phi_+\right>=\left<\phi_-\right>=\left<\phi_0\right>=0$, 
supersymmetry is unbroken if the potential vanishes in the
minimum. As can be seen from the potential \eqref{V} this implies that 
\begin{equation}
\left<D^3(y)\left>-\pa_y\right<\Phi(y)\right>= 0
\end{equation}
for a supersymmetric background field configuration. As this is a
first order differential equation for the odd $\left< \Phi \right>$,  
which vanishes at the boundaries, a supersymmetric VEV is only
possible if the constant mode $D^3_0 = \int dy\, D^3(y)$ of the
auxiliary field \eqref{D3}
\begin{equation}
D^3_0 
=
\frac{g}{\pi R}
\!\left(\!
(q+q_0)\frac{\La^2}{16\pi^2}
+q_0\left|\left<\phi_0\right>\right|^2
+q\int_0^{\pi R}\dd y
\left(\left|\left<\phi_+(y)\right>\right|^2
-\left|\left<\phi_-(y)\right>\right|^2\right)
\!\right)\!
\label{D30}
\end{equation}
vanishes. (The logarithmic divergence is absent here, since the
derivative $\delta'(y)$ vanishes at $y = 0$.) 
Hence, as far as the auxiliary field $D_3$ is concerned, 
only its zero mode $D^3_0$ is important for
determining whether supersymmetric minimum can exist, 
in the presence of FI--terms localized at the branes in the 
5--dimensional theory.  
%
%

Let us consider two cases: {\it i)} the charges of the matter fields
do not sum up to zero, $q+q_0\ne0$;
{\it ii)} the charges sum up to zero, $q+q_0=0$.
These cases are different because the radiatively generated
quadratically divergent contribution to $D^3_0$ is proportional to
the sum of charges $q+q_0$ (see (\ref{D30})).

In case {\it i)} the
supersymmetry can remain unbroken only when some nonzero VEVs of the
charged fields cancel that $\La^2$ contribution to $D^3_0$:
\begin{equation}
(q+q_0)\frac{\La^2}{16\pi^2}
+q_0\left|\left<\phi_0\right>\right|^2
+q\int_0^{\pi R}\dd y
\left(\left|\left<\phi_+(y)\right>\right|^2
-\left|\left<\phi_-(y)\right>\right|^2\right)
=
0
\,.
\label{SUSY}
\end{equation}
The bulk field $\phi_0$ can be used for this purpose if its charge
$q_0$ has opposite sign with respect to the sum $q+q_0$ 
(i.e.\ $qq_0<0$ and $|q_0|<|q|$).

The situation with the bulk field is more complicated. 
We have to remember that its VEV must minimize the whole potential 
$V$ and not only its first term depending on
$D^3$. Only one of the components, $\phi_+$ or $\phi_-$, can have
nonvanishing VEV because of the $|\phi_+\phi_-|^2$ term in the
potential (\ref{V}). From the last two terms in $V$ we 
see that such nonzero VEVs must satisfy 
\begin{equation}
\pa_y\left<\phi_\pm(y)\right>
=
\mp gq\left<\Phi(y)\right>\left<\phi_\pm(y)\right>
\,.
\label{phiVEV}
\end{equation}
In the case of $\left<\phi_+\right>$ it can be solved 
with the normalization chosen to fulfill eq.\ (\ref{SUSY}). 
The configuration with such $\left<\phi_+\right>$ and 
$\left<\phi_-\right>=0=\left<\phi_0\right>$ gives vanishing value of
$V$ which is a global minimum because $V$ is positive
semi--definite.
Nonzero $\left<\phi_+\right>$ can be used to restore supersymmetry
if $q$ has opposite sign with respect to the sum $q+q_0$ 
($qq_0<0$ and $|q|<|q_0|$).

Equation (\ref{phiVEV}) has no nonzero solutions for the odd field
$\phi_-$. The reason is that for $\lim_{y\to 0}\phi_-(y)\ne 0$ 
the l.h.s.\ of (\ref{phiVEV}) has a
$\de(y)$ type singularity while no such structure appears
at the r.h.s.\ of that equation. As the result $\left<\phi_-\right>$
vanishes and can not cancel the $\La^2$ term in $D^3_0$.

We see that there are two possibilities in the case {\it i)}
($q+q_0\ne0$). 
If the bulk and brane field charges have opposite signs then one of
these fields (the one with the smaller absolute value of the charge) 
develops a nonzero VEV. This VEV is such that the vacuum
energy vanishes and the supersymmetry remains unbroken. 
But the $U(1)$ gauge symmetry is broken and the scale of breaking is
given by the cut off scale $\La$. If $q$ and $q_0$ have the same sign
then $U(1)$ is unbroken because the charged fields do not develop
nonzero VEVs. 
(As argued above, the oppositely charged odd field $\phi_-$ cannot
have a VEV.)
In such a case the value of the vacuum energy is
positive and supersymmetry is broken at 
very high scale given by $\La$.
Thus the situation is very similar to the 4--dimensional theories with
quadratically divergent Fayet--Iliopoulos terms generated radiatively
if the $U(1)$ charges do not sum up to zero.

Let us now go to the case {\it ii)} in which charges do sum up to zero:
$q+q_0=0$. Now there is no $\La^2$ contribution to $D^3_0$ given by
(\ref{D30}). It is obvious that the potential $V$ is minimized
for vanishing VEVs of all matter fields. The gauge $U(1)$ symmetry and
the supersymmetry are unbroken. This is again very similar to the
4--dimensional theory but there are important differences: In 4
dimensions no FI--term is generated in such a case. In our bulk--brane
setup nontrivial FI--like term is generated for the combination
$(D^3-\pa_y\Phi)$ and only its integral over $y$ vanishes. 
The supersymmetry remains unbroken because the
contributions to the VEV of the auxiliary field coming from $D^3$ and
$\pa_y\Phi$ exactly cancel each other (this is related to the fact
that $\Phi$ is a propagating field while $D^3$ is not). 
But we will see that the non-trivial FI--term for the propagating field
$\Phi$ has very interesting consequences. 
%
%

First we calculate the background configuration 
of $\Phi$ which is induced by the FI--terms.  
Using equations (\ref{xi}), (\ref{V}) and (\ref{D3}) we get for
$q_0=-q$:
\begin{equation}
\left<\Phi(y)\right>=
-gq\frac{\La^2}{64\pi^2}\eps(y)
+gq\frac{\ln\La^2}{32\pi^2}\left(\de'(y)+\de'(y-\pi R)\right)
\label{PhiVEV0}
\end{equation}
where we again drop terms which are not sensitive to the scale $\La$.
Substituting this back to the part of the potential (\ref{V})
\begin{equation}
\int\dd y
\big(
\left|\pa_y\phi_+-gq\left<\Phi\right>\phi_+\right|^2
+\left|\pa_y\phi_-+gq\left<\Phi\right>\phi_-\right|^2
\big)
\end{equation}
gives the mass terms for the fields $\phi_+$ and $\phi_-$ in the
effective 4--dimensional theory. The theory is supersymmetric so the
nonzero VEV of $\Phi$ influences also the effective mass term for the
matter fermions (via the last term in (\ref{Lbulk})) 
\begin{equation}
\int\dd y
\bar\psi\left(i\ga^5\pa_y+gq\left<\Phi\right>\right)\psi
\,.
\end{equation}

With vanishing $\left<\Phi\right>$ the spectrum of $\phi_\pm$ is very
simple: The even component 
has a constant massless mode while modes $\phi_+(y)=\cos(ny/R)$,
$\phi_-(y)=\sin(ny/R)$ have masses equal to $n/R$. Let us now analyse
the spectrum of the charged bulk fields with VEV of $\Phi$ given by
(\ref{PhiVEV0}). First of all the even field $\phi_+$ has again a zero
mode given formally by the formula
\begin{equation}
\phi_{+0}(y)
=
\exp\left(gq\int_0^y\dd y'\left<\Phi(y')\right>\right)\phi_{+0}(0)
\,.
\label{phi+0}
\end{equation}
We can not directly substitute here $\left<\Phi\right>$ as given by
(\ref{PhiVEV0}) because we would get arbitrary powers of $\de'(y)$
which have no well definite meaning. We regularize the delta function
with the appropriate Gaussian function so that
\begin{equation}
\de'(y)
=
-\lim_{\si\to\infty}2\pi\si^3y\exp\left(-\pi\si^2y^2\right)
\,.
\end{equation}
Now the integral (\ref{phi+0}) can be calculated
\begin{eqnarray}
\phi_{+0}(y)
=
N\exp\Big\{\!
-\frac{g^2q^2}{64\pi^2}
\La^2y\Big\}
\exp\Big\{\!
-
\frac{g^2q^2}{32\pi^2}
\ln(\La)\si\Big[
1
\!\!\!&+&\!\!\!
e^{-\pi^2\si^2}-
e^{-\pi\si^2y^2}
\nn\\
\!\!\!&-&\!\!\!
e^{-\pi\si^2(y-\pi R)^2}
\Big]
\Big\}.
\quad
\end{eqnarray}
$N$ is a normalization factor depending on $\La$ and $\si$.
The expression in the second curly bracket is negative for 
$0<y<\pi R$ and vanishes only at the fixed points. 
Thus the second exponential factor suppresses 
the zero mode very strongly for all points which are not close to 
$y=0$ or $y=\pi R$. The first exponential factor involving $\La^2$
suppresses further the zero mode for $y$ away from 0. 
As a result the zero mode is 
exponentially localized at the $y=0$ brane and the width of this
localization is determined by the bigger of the scales $\La$ and
$\si$. The cut off $\La$ can be treated as a scale characteristic 
for some more fundamental underlying theory, e.g. the string theory.
The scale $\si$ was introduced in order to resolve the infinitely thin
branes. The thickness of the branes is likely to be determined by the
scale characteristic for a theory describing brane dynamics (it could
be related to $\La$ in a more fundamental theory). 
Anyway, the localization of the zero mode
of $\phi_+$ is not weaker than the localization of the brane itself
(parameterized by $\si$). The FI--terms cause the zero mode
$\phi_{+0}$ to become effectively a brane field localized at the same
brane at which the oppositely charged field $\phi_0$ is localized. 
Of course the corresponding fermionic zero mode of $\psi$ is also
localized at the same brane. This has important implications for the
anomaly analysis in such models. For the case $q_0=-q$ the anomaly is
canceled locally at this brane (and is absent at the other brane).

We will not discuss in much detail the massive modes of the bulk
fields. The important point is that they effectively disappear from
the spectrum because their masses are at least as big as the cut off
scale $\La$. This can be seen by analyzing first the spectrum without 
the $\ln\La$ term in (\ref{PhiVEV0}). In such a case the mass
eigenstates are given by 
\begin{equation}
\left(
\begin{array}{c}
\phi_{+n}
\\
\phi_{-n}
\end{array}
\right)
=
\sqrt{\frac{2}{\pi R}}
\left(
\begin{array}{l}
\cos\left[ny/R-\al\right]
\\
\sin\left[ny/R\right]
\end{array}
\right)
\end{equation}
where the phase $\al$ is given by $\tan\al=gqR\La^2/64\pi^2n$
and the corresponding masses are 
$m_n^2=(gqR\La^2/64\pi^2)^2+(n/R)^2$. Then one can show that adding 
to $\left<\Phi\right>$  a term depending on $\si$ changes the profiles
of the eigenstates but do not change their masses (because this new
term integrates to zero and thus do not change the boundary conditions 
which are used to get the discrete spectrum).

A similar localization of the bulk fields takes place also 
when $q+q_0\ne0$. However the formulae in such a case are more
complicated and will be discussed elsewhere \cite{future}.

We showed that the bulk matter field get localized at the brane due to
the divergent FI--terms. One could ask the question whether such terms
cause also the bulk gauge field to localize? The answer is: no. The
gauge field have no self-couplings (it is an Abelian gauge symmetry)
hence VEV of $\Phi$ does not change the ordinary Kaluza--Klein tower
of states coming from the bulk gauge multiplet.

In the present paper we have 
analyzed the 5--dimensional model compactified on $S^1/\ZZ_2$ with
one brane and one bulk matter multiplets coupled to the bulk $U(1)$
gauge multiplet. It occurs that the Fayet--Iliopoulos terms are
generated not only for the auxiliary component of the gauge multiplet,
$D^3$, but also for the derivative of the scalar component,
$\pa_y\Phi$. The potential with these FI--terms added has been analyzed
in order to check whether supersymmetry or gauge symmetry is
spontaneously broken. The situation is very similar to the
4--dimensional case. Both supersymmetry and gauge symmetry remains
unbroken only if the charges of matter fields sum up to zero. If the
charges do not sum up to zero one of these symmetries is
broken. Supersymmetry is broken if all the charges have the same sign.
If charges of both signs exist then one of the matter fields 
develops a nonzero VEV and the gauge symmetry is broken. The scale of
the breaking in both cases is of the order of the cut off scale $\La$
which should be regarded as a scale characteristic for an underlying
more fundamental theory.

As a main result of this paper we have shown
that the FI--terms have another very interesting
consequences for this bulk--brane model. Namely, they lead to the
spontaneous localization of the charged bulk matter field. The zero
mode of the bulk field is localized at the brane at which the bulk
field lives (if their charges have opposite signs). The width of such
localization is related to the thickness of the brane. The massive
modes of the bulk field get masses at least of the order of the cut
off scale and disappear effectively from the spectrum. Thus, the bulk
matter field changes spontaneously into a brane field. 
It seems that the model with the single bulk charged matter field
develops this kind if instability.

We have discussed the simplest model with just one brane matter field, 
one bulk matter field and one bulk gauge field. However, the main
features are very similar for more complicated models in which the
$U(1)$ symmetry is just the hypercharge part of the Standard Model
gauge group and there are many bulk and brane matter fields. 
We will consider more general models and their phenomenological
applications in a future publication \cite{future}. 
There we will also discuss in more detail consequences of this
localization for the anomaly analysis.


\section*{Acknowledgments}
Work supported in part by the European Community's Human Potential
Programme under contracts HPRN--CT--2000--00131 Quantum Spacetime,
HPRN--CT--2000--00148 Physics Across the Present Energy Frontier
and HPRN--CT--2000--00152 Supersymmetry and the Early Universe.
SGN was supported by priority grant 1096 of the Deutsche
Forschungsgemeinschaft. 
MO was partially supported by the Polish KBN grant 2 P03B 052 16.

\end{document}